\documentclass[12pt]{article}
\usepackage{graphicx}
\usepackage{epstopdf}
\usepackage{caption}
\usepackage{subcaption}
\usepackage{subfig}

\usepackage{epsfig,amsmath,amssymb,verbatim,mathrsfs}

\def\beq{\begin{eqnarray}}
\def\eeq{\end{eqnarray}}
\def\bea{\begin{eqnarray}}
\def\eea{\end{eqnarray}}

\newcommand{\h}{\text{\tiny H}}
\newcommand{\sm}{\text{\tiny SM}}
\newcommand{\sh}{\text{\tiny SH}}
\newcommand{\dm}{\text{\tiny DM}}
\newcommand{\cdm}{\text{\tiny CDM}}

\begin{document}

\setlength{\baselineskip}{0.2in}


\begin{titlepage}
\noindent
\flushright{May 2013}
\vspace{0.2cm}

\begin{center}
  \begin{Large}
    \begin{bf}
A Class of Inert N-tuplet Models with Radiative Neutrino Mass and Dark Matter\\

     \end{bf}
  \end{Large}
\end{center}

\vspace{0.2cm}

\begin{center}

\begin{large}

{Sandy~S.~C.~Law$^{*}$\footnote{slaw@mail.ncku.edu.tw}  and Kristian~L.~McDonald$^{\dagger}$\footnote{klmcd@physics.usyd.edu.au}}\\
     \end{large}
\vspace{0.5cm}
  \begin{it}
* Department of Physics, National Cheng-Kung University, \\ Tainan 701, Taiwan\\
\vspace{0.5cm}
$\dagger$ ARC Centre of Excellence for Particle Physics at the Terascale,\\
School of Physics, The University of Sydney, NSW 2006, Australia\\\vspace{0.5cm}
\vspace{0.3cm}
\end{it}
\vspace{0.5cm}

\end{center}


\begin{abstract}
We present a class of models with radiative neutrino mass and stable dark-matter candidates. Neutrino mass is generated by a one-loop diagram with the same topography as Ma's 2006 proposal (which used an inert scalar-doublet and singlet fermion). We generalize this approach and determine all variants with new fields no larger than the adjoint representation.  When the neutrino mass diagram contains a Majorana mass insertion there are two possibilities, both of which are known. If the mass insertion is of the Dirac type there are seven additional models, two of which are excluded by direct-detection experiments. The  other five models are also constrained, such that only scalar dark-matter is viable. There are cases with an inert singlet, an inert doublet, and an inert triplet, providing a natural setting for inert $N$-tuplet models of dark matter, with the additional feature of achieving radiative neutrino mass. We  show that some of the  models admit a simple explanation for the (requisite) discrete symmetry, and briefly  discuss cases with representations larger than the adjoint, which can admit a connection to the astrophysical gamma-ray signal.

\end{abstract}

\vspace{1cm}

\end{titlepage}

\setcounter{page}{1}


\vfill\eject


\section{Introduction\label{sec:introduction}}

The experimental evidence for neutrino mass, acquired in recent decades, provides concrete evidence for physics beyond the Standard Model (SM) (see e.g.~\cite{GonzalezGarcia:2012sz}). Although the requisite new degrees of freedom cannot yet be determined it is clear that additional particles are likely to exist, in order to generate the masses. Similarly, there is by now a large amount of evidence for an additional galactic constituent, an unknown substance referred to as dark matter (see e.g.~\cite{Peter:2012rz}). This may or may not require new degree's of freedom, but the hypothesis that the dark matter is comprised of a stable (or long lived) new particle species provides a simple explanation for this observed feature of the Universe. Given that these two indicators for  beyond-SM physics can be explained by extending the particle spectrum of the SM, it is natural to ask if the requisite new particles can be related. Could the mechanism of neutrino mass be related to the existence of a stable dark-matter candidate?

A particularly simple model realizing this idea was proposed by Ma in 2006~\cite{Ma:2006km}. This model extends the SM to include an additional SM-like scalar doublet and gauge-singlet fermions, all of which are odd under a discrete $Z_2$ symmetry. The extended field content allows for radiative neutrino mass, generated at the one-loop level, while the lightest beyond-SM field is absolutely stable. One thus arrives at a simple synergetic model of radiative neutrino mass and dark matter.

\begin{figure}[ttt]
\begin{center}
        \includegraphics[width = 0.6\textwidth]{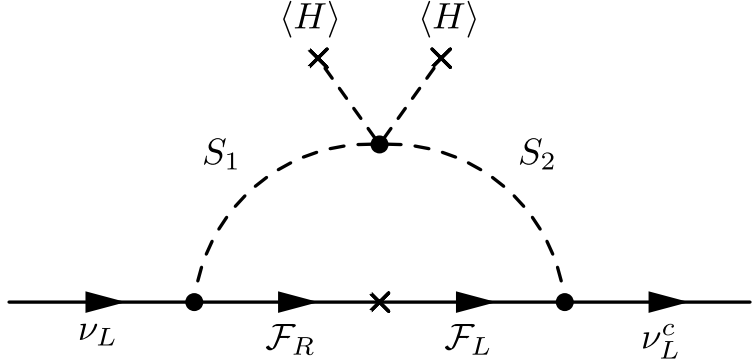}
\end{center}
\caption{The generic one-loop diagram for neutrino mass in a class of models that admit stable dark-matter candidates. Here $\mathcal{F}$ and $S_{1,2}$ are beyond-SM multiplets that collectively contain at least one electrically-neutral field.}\label{fig:yukawa_loop}
\end{figure}

In this work we generalize Ma's approach. We present a class of related models, all of which generate neutrino mass via a loop diagram with the same topography as Ma's, whilst simultaneously admitting stable dark-matter candidates. The loop-diagram employed by Ma contains a mass insertion on the internal fermion line (see Figure~\ref{fig:yukawa_loop}), and our generalizations fall naturally into two categories; those which break lepton-number symmetry via a Majorana mass insertion, and those with a Dirac mass insertion, such that lepton-number symmetry is broken at a vertex. Although the basic mechanism is very similar in both cases, this difference modifies one's expectations for the beyond-SM field content and the associated phenomenology.

It turns out that, in both cases, this approach is very general and many realizations are possible. However, restricting attention to models in which the beyond-SM multiplets are no larger than the adjoint representation significantly  reduces the possibilities. As we shall see, there are only two such (minimal) models with a Majorana mass-insertion, both of which are known~\cite{Ma:2006km,Ma:2008cu}.  We find seven additional models with a mass insertion of the Dirac type, all of which achieve radiative neutrino mass and dark-matter candidates. We detail these models, finding a subset that are compatible with direct-detection experiments. There are cases with an inert singlet, an inert doublet, and an inert triplet; the models therefore provide a natural setting for inert $N$-tuplet theories of dark matter, such that radiative neutrino mass is also achieved. 

Interestingly, three of the new models admit a simple extension that can explain the origin of the  (formerly imposed)  discrete symmetry. By upgrading the discrete symmetry to a gauged $U(1)$ symmetry, and extending the field content by a single SM-singlet scalar, the discrete symmetry can arise as an accidental symmetry of the low-energy Lagrangian, after $U(1)$ symmetry breaking takes place. This provides a simple explanation for the discrete symmetry.

Though we focus on models with representations no larger than the adjoint, we also briefly discuss cases where the beyond-SM fields can be quadruplet and/or quintuplet representations of $SU(2)_L$. We present the candidate models in these cases, and mention some key issues, based on the lessons learned from our preceding studies. Despite the use of larger multiplets, these models can still be of interest; in addition to allowing for radiative neutrino mass, the exotics with larger electric-charges in these multiplets can enhance the $2\gamma$ and/or $\gamma+Z$ signal from dark-matter annihilation when they appear inside loops~\cite{Kopp:2013mi}. This can provide a simple connection between the mechanism of neutrino mass and the astrophysical gamma-ray signal~\cite{Weniger:2012tx}.

Before proceeding we note that the connection between radiative neutrino mass and dark matter has been explored in a number of different models, including Ref.~\cite{Krauss:2002px}, which precedes Ma's work; for other examples see~\cite{Boehm:2006mi}. Previous works on inert-singlet models~\cite{McDonald:1993ex}, inert-doublet models~\cite{Barbieri:2006dq}, and inert-triplet models~\cite{Cirelli:2005uq} are also well known. Additional relevant works dealing with inert-multiplet dark matter and/or radiative neutrino mass are cited in the text. Note also that Refs.~\cite{Ma:1998dn,Bonnet:2012kz} have detailed the one-loop realizations of the $d=5$ operator for neutrino mass. Inert scalar dark matter can also help cure the little hierarchy problem found in low-scale seesaws~\cite{Fabbrichesi:2013qca}. The present work follows on from the generalized tree-level seesaws presented in Ref.~\cite{McDonald:2013kca}.

The plan of this paper is as follows. In Section~\ref{sec:rad_nu_plus_dm} we discuss Ma's model and present the generalizations that similarly achieve radiative neutrino mass and dark-matter candidates. Section~\ref{sec:fermion_dm} considers the case of fermionic dark-matter, while Section~\ref{sec:scalar_dm} discusses scalar dark-matter. One of the generalized models is presented in more detail in Section~\ref{sec:real_and_complex_scalar}.   In Section~\ref{sec:gauge_symmetry} we show that some of the models allow a simple extension, such that the discrete symmetry appears as an accidental symmetry in the low-energy theory.  Models with exotics forming larger $SU(2)$ representations are discussed in Section~\ref{sec:non-minimal_models} (and explicitly displayed in the Appendix). We conclude in Section~\ref{sec:conc}.

\section{Radiative Neutrino Mass and Dark Matter\label{sec:rad_nu_plus_dm}}
We are interested in the class of models that generate neutrino mass radiatively by the diagram in Figure~\ref{fig:yukawa_loop}. Here $\mathcal{F}$ is a beyond-SM fermion and $S_{1,2}$ are new scalars (which can be identical in some cases). A basic feature of this diagram is that the three vertices can all involve two beyond-SM fields. Consequently one can always consider a discrete $Z_2$ symmetry whose action on the beyond-SM multiplets is
\bea
\{\mathcal{F},\; S_1,\; S_2\}&\rightarrow& - \ \{\mathcal{F},\; S_1,\; S_2\}\;,
\eea
while all SM fields transform trivially. The lightest field within the multiplets $\mathcal{F}$ and $S_{1,2}$ will thus be stable, and provided this field is electrically neutral and colorless, one arrives at a dark-matter candidate. These comments are generic for all models of this type; the connection between loop masses and dark matter is simple to realize in this class of models.

Figure~\ref{fig:yukawa_loop} produces Majorana neutrino masses, so the loop diagram must contain a source of lepton number violation. Choosing the convention for lepton-number symmetry such that the new fermion $\mathcal{F}$ has the same value as the SM leptons, there are two ways to explicitly break lepton-number symmetry; it can be broken  at either the mass insertion or at one of the vertices. The simplest models, in terms of the requisite number of beyond-SM multiplets, are those with a lepton number violating (Majorana) mass insertion. In this case one has $\mathcal{F}_L\equiv \mathcal{F}_R^c$ and minimal cases occur for $S_1=S_2\equiv S$. Thus, only two beyond-SM multiplets are required. The general loop-diagram for this subset of models is given  in Figure~\ref{fig:mass_loop}. We consider this case first.
\subsection{Models with a Majorana Mass Insertion\label{sec:maj_mass}}
\begin{figure}[ttt]
\begin{center}
        \includegraphics[width = 0.6\textwidth]{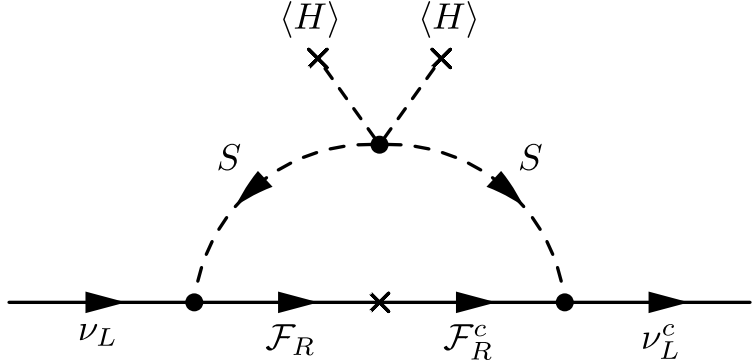}
\end{center}
\caption{The generic one-loop diagram for models with a lepton number violating mass insertion.}\label{fig:mass_loop}
\end{figure}

We seek models that achieve neutrino mass via Figure~\ref{fig:mass_loop} and give rise to dark-matter candidates. Clearly the fermion must form a real representation of the SM gauge symmetry, $\mathcal{F}_R\sim(1,R_{\mathcal{F}},0)$, in order to allow a bare Majorana mass. As we are considering dark-matter candidates we do not consider colored fields. Note also that $R_{\mathcal{F}}$ must be odd-valued to ensure there is no fractionally charged particles (the lightest of which would be stable and thus cosmologically excluded). Odd-valued $R_{\mathcal{F}}$ also ensures that $\mathcal{F}_R$ contains an electrically neutral component, so no additional constraint is imposed by this demand.

With this information one can obtain the viable combinations of $\mathcal{F}_R$ and $S$ that generate Figure~\ref{fig:mass_loop}. The basic Lagrangian terms are
\bea
\mathcal{L}&\supset& i\bar{\mathcal{F}_R}\gamma^\mu D_\mu\mathcal{F}_R - \frac{M_{\mathcal{F}}}{2}\;\overline{\mathcal{F}_R^c}\mathcal{F}_R +|D^\mu S|^2 -M_S^2 |S|^2 +\lambda\bar{L} \tilde{S}\mathcal{F}_R+\lambda_{\sh}(S^\dagger H)^2 +\mathrm{H.c.},\label{eq:ma_lagrangian}
\eea
where $L$ ($H$) is the SM lepton (scalar) doublet and $\tilde{S}$ denotes the charge-conjugate of $S$. It turns out that the possible combinations for $\mathcal{F}$ and $S$ are not restricted by quantum numbers; one can consider increasingly large multiplets, presumably up to some unitarity limits~\cite{Hally:2012pu}, and realize a model with Figure~\ref{fig:mass_loop} and a dark-matter candidate. However, if we restrict our attention to models with $R_{\mathcal{F}},R_S\le 3$, such that no new multiplet is larger than the adjoint representation, there are only two possibilities. The first case is Ma's original proposal, which employs an additional (inert) scalar doublet $S\sim (1,2,1)$, and a gauge-singlet fermion $\mathcal{F}_R\sim(1,1,0)$~\cite{Ma:2006km}. This model is the prototype for the class we are considering. The second model also employs the scalar doublet $S\sim (1,2,1)$, but instead utilizes the triplet fermion $\mathcal{F}_R\sim(1,3,0)$~\cite{Ma:2008cu}, familiar from the Type-III seesaw~\cite{Foot:1988aq}. Thus, both of the models are known in the literature, and there are no additional possibilities unless one considers larger multiplets.

In each of these models the dark-matter candidate can be a neutral component of $S$ or a Majorana fermion. There is, however,  an important difference between the singlet case, $\mathcal{F}\sim(1,1,0)$, and the other model; the singlet does not participate in weak interactions and is therefore brought into thermal contact with the SM sector via the Yukawa coupling. For fermionic dark-matter, this can produce conflict between the need to keep the Yukawa coupling large to ensure thermal dark-matter, and the need to suppress the Yukawa coupling to limit the size of flavor changing effects. This issue does not arise in the triplet fermion model, as the fermions can maintain equilibrium with the SM sector via weak interactions in these cases, even if the Yukawa couplings are suppressed. For an analysis of Ma's model, incorporating recent LHC data on the Higgs, see e.g.~Ref.~\cite{Ho:2013hia}. Also note that loop effects can induce observable interactions between dark matter and experimental detectors in Ma's model~\cite{Schmidt:2012yg}.

\subsection{Models with a Dirac Mass Insertion\label{sec:dirac_mass}}

Having exhausted the minimal models with a lepton-number violating mass insertion, we now consider models with a Dirac mass insertion; i.e.~the beyond-SM fermion has nonzero hypercharge. In this case the general mass-diagram has the form shown in Figure~\ref{fig:yukawa_loop}. The fields $\mathcal{F}_R$ and $\mathcal{F}_L$ are no longer related by charge conjugation, so the mass insertion is of the Dirac type and $\mathcal{F}$ is a vector-like fermion. In addition one requires $S_1\ne S_2$. We again consider a $Z_2$ symmetry under which the SM fields transform trivially but the new fields are odd. The Lagrangian contains the following pertinent terms:
\bea
\mathcal{L}&\supset& i\bar{\mathcal{F}}\gamma^\mu D_\mu\mathcal{F}\ -\  M_{\mathcal{F}}\;\overline{\mathcal{F}}\mathcal{F} \ +\ \sum_{i=1,2}\left\{|D^\mu S_i|^2 \ -\ M_i^2 |S_i|^2 \right\}\nonumber\\
& & +\ \lambda_1\;\bar{L} \mathcal{F}_RS_1\ +\ \lambda_2\; \bar{L}\mathcal{F}^c_L\tilde{S}_2\ +\ \lambda_{\sh}\;S_1\tilde{S}_2 H^2 +\mathrm{H.c.},\label{eq:L_ykawa_lagrange}
\eea
where, in our convention, lepton number symmetry is broken by a Yukawa coupling. With the $Z_2$ symmetry present, there are no terms  in the scalar potential that are linear in \emph{just one} of the new scalars $S_{1,2}$. Therefore the beyond-SM scalars do not acquire an induced VEV and parameter space exists for which $\langle S_{1,2}\rangle=0$, so the $Z_2$ symmetry remains exact.

In selecting viable multiplets one must ensure that no new multiplet contains a fractionally charged field, to avoid  a (cosmologically excluded) stable charged field. To ensure that the lightest $Z_2$-odd field is a neutral dark-matter candidate  one must demand that the new multiplets contain at least one neutral field. Note that the neutral field does not have to appear as an explicit propagating degree of freedom inside the loop diagram; it is sufficient merely that the loop-diagram exists and that the particle content includes a neutral field that can play the role of dark matter.

With these conditions in mind we search for viable combinations of the beyond-SM multiplets that realize Figure~\ref{fig:yukawa_loop}. We find that the size of the beyond-SM multiplets is not restricted by our demands; one can consider increasingly large multiplets and realize the loop diagram. However, if attention is restricted to models in which none of the beyond-SM multiplets are larger than the  adjoint representation, only seven distinct models are found. These are listed in Table~\ref{L_yukawa}. Of the seven  models, one employs the exotic lepton triplet $\mathcal{F}\sim(1,3,-2)$, studied in Refs.~\cite{Chua:2010me,DelNobile:2009st}, three contain an exotic vector-like (and SM-like) lepton doublet $\mathcal{F}\sim(1,2,-1)$~\cite{DelNobile:2009st,Joglekar:2012vc,Arina:2012aj}, and two contain a charged lepton doublet $\mathcal{F}\sim(1,2,-3)$~\cite{DelNobile:2009st,Law:2011qe}. There is also a model with a SM-like charged singlet fermion, $\mathcal{F}\sim(1,1,-2)$, which already appeared in Ref.~\cite{Aoki:2011yk}.

Neutrino masses take a standard calculable form in these models. For example, in models $(A)$ and $(B)$ only singly-charged exotics propagate in the loop, and the SM neutrino mass matrix is given by\footnote{The quoted result is for model $(B)$, and  should be multiplied by an extra factor of 2 for model $(A)$.}
\bea
(\mathcal{M}_\nu)_{\alpha\beta}&\simeq& \frac{\left[(\lambda_2^*)^a_{\alpha}(\lambda_1^*)^a_{\beta}+(\lambda_2^*)^a_{\beta}(\lambda_1^*)^a_{\alpha}\right]}{32\pi^2}\,\frac{\lambda_{\sh}\langle H\rangle^2}{M_{>}^2-M_{<}^2}\left[ \frac{M^2_{>}\,M_{\mathcal{F},a}}{M_{\mathcal{F},a}^2-M_>^2}\,\log \frac{M_{\mathcal{F},a}^2}{M_{>}^2}\ -\ (M_>\rightarrow M_<)\right]\nonumber\\
& &
\eea
Here $M_{\mathcal{F},a}$ is the mass for the charged component of the exotic fermion $\mathcal{F}_a$, and summation is implied for the repeated index $a$ (which labels the exotic-fermion generations).\footnote{A single generation of exotic fermions generates nonzero  masses for two SM neutrinos, which  is sufficient to accommodate the experimentally observed mass and mixing spectrum. To obtain three massive neutrinos requires (at least) two generations of exotic fermions.} The masses $M_{>,<}$ refer to the charged scalar mass-eigenstates, which are linear combinations of the charged scalars $S_1^+$ and $S_2^+$. The mixing results from the $\lambda_{\sh}$-term in Eq.~\eqref{eq:L_ykawa_lagrange}, which takes the explicit form $\lambda_{\sh} \tilde{H}^\dagger S_1 S_2^\dagger H\subset\mathcal{L}$, for models $(A)$ and $(B)$.  If all the exotics are at the TeV scale one requires dimensionless couplings of $\mathcal{O}(10^{-3})$ to obtain $m_\nu\sim0.1$~eV. The scenario with all exotics at the TeV scale is most interesting from a phenomenological perspective. However, strictly speaking one only requires the lightest exotic to have a mass of $\lesssim\mathcal{O}(\mathrm{TeV})$ in order to realize a dark-matter candidate. The other exotics can be much heavier, allowing larger dimensionless couplings.

Before moving on to discuss dark matter in detail, we note that, of the models in Table~\ref{L_yukawa}, only models $(A)$, $(B)$ and $(D)$ are expected to produce (dominant) radiative neutrino masses in the absence of the $Z_2$ symmetry, as we briefly discuss in Appendix~\ref{app:mass_no_Z2}. 


\begin{table}
\centering
\begin{tabular}{|c|c|c|c|c|}\hline
& & &  &\\
\ \ Model\ \ &
$\mathcal{F}$  & $S_1$&$S_2$&Dark Matter Status\\
& & &  &\\
\hline
& & &  &\\
$(A)$& $\ \ (1,1,-2)\ \ $ &$\ \ (1,2,1)\ \ $&\ \ $(1,2,3)$\ \ &Inert Doublet  (Ref.~\cite{Aoki:2011yk})\\ 
& & &  &\\
\hline
& & &  &\\
$(B)$& $\ \ (1,3,-2)\ \ $ &$\ \ (1,2,1)\ \ $&\ \ $(1,2,3)$\ \ &Inert Doublet \\ 
& & &  &\\
\hline
& & &  &\\
$(C)$& $(1,2,-1)$ &$(1,1,0)$&\ $(1,3,2)$\ &Inert Singlet or Triplet\\ 
& & &  &\\
\hline
& & &  &\\
$(D)$& $(1,2,-1)$ &$(1,3,0)$&$(1,1,2)$&Inert Real Triplet \\ 
& & &  &\\
\hline
& & &  &\\
$(E)$& $(1,2,-1)$ &$(1,3,0)$ &$(1,3,2)$&Inert Triplet\\ 
& & &  &\\
\hline
& & &  &\\
$(F)$& $(1,2,-3)$ &$(1,3,2)$&$(1,1,4)$&Excluded (Direct Detection)\\ 
& & &  &\\
\hline
& & &  &\\
$(G)$& $(1,2,-3)$ &$(1,3,2)$ &$(1,3,4)$&Excluded (Direct Detection)\\ 
& & &  &\\
\hline
\end{tabular}
\caption{\label{L_yukawa} Minimal models with a Dirac mass insertion. These models allow radiative neutrino mass and contain at least one stable  electrically-neutral beyond-SM field.}
\end{table}

\section{Inert Fermionic Dark Matter\label{sec:fermion_dm}}
We now turn our attention to the dark-matter candidates in these models. It is \emph{a priori} possible that both fermionic and scalar dark-matter candidates are possible, as in Ma's original proposal~\cite{Ma:2006km}. In this section we consider fermionic dark-matter. Note that not all the models contain neutral beyond-SM fermions; specifically, model $(A)$ has $\mathcal{F}\sim(1,1,-2)$, and models $(F)$ and $(G)$ use $\mathcal{F}\sim(1,2,-3)$.  In these cases all beyond-SM fermions are charged and only scalar dark-matter is possible. One can already exclude  the parameter space with light fermions in these models, namely $M_{\mathcal{F}}<M_{1,2}$, due to the appearance of a stable charged-fermion.

On the other hand, models $(B)$ through $(E)$ all contain neutral fermions and can, in principle, admit fermionic dark-matter. However, all of the fermion multiplets in these models have nonzero hypercharge, which can lead to strong constraints from direct-detection experiments. More precisely, if the dark-matter abundance is generated by a standard thermal WIMP one can exclude Dirac-fermion dark-matter with nonzero hypercharge, due to the strong constraints from e.g.  XENON100~\cite{Aprile:2011hi}. Thus, it is important to determine whether the neutral fermion is Dirac or Majorana.  

\begin{figure}[ttt]
\begin{center}
        \includegraphics[width = 0.7\textwidth]{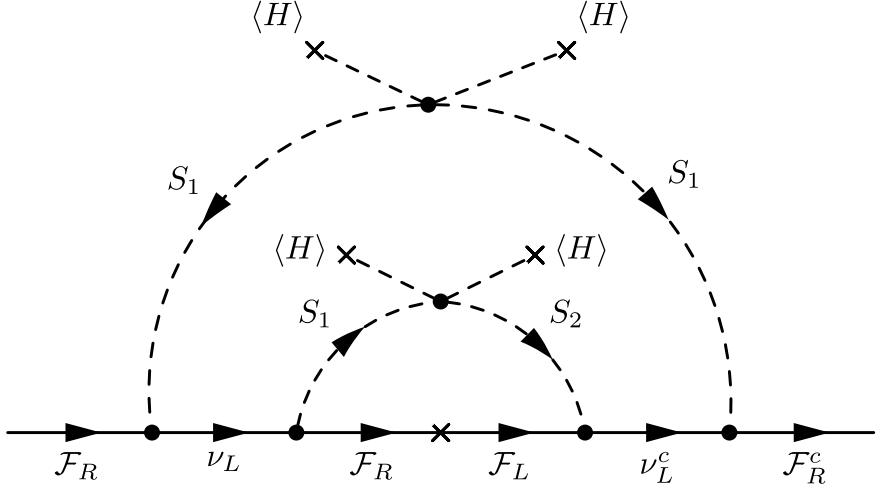}
\end{center}
\caption{A loop diagram that generates a Majorana mass for the fermion $\mathcal{F}_R\sim(1,3,-2)$. Here $S_1\sim(1,2,1)$ and $S_2\sim(1,2,3)$ [model $(B)$]. Similar diagrams exist for models $(C)$ through $(E)$.}\label{fig:F_maj_mass}
\end{figure}

At tree-level the fermion $\mathcal{F}$ remains a Dirac particle. However, its coupling to the SM neutrinos, which obtain Majorana masses via Figure~\ref{fig:yukawa_loop}, leads to a small radiative Majorana-mass. For the case of $\mathcal{F}\sim(1,3,-2)$, the typical diagram is shown in Figure~\ref{fig:F_maj_mass}.  Similar diagrams can occur for models $(C)$ through $(E)$, though the scalar $S_1$ is real in these cases. The loop-induced Majorana mass will, in general, split the Dirac fermion $\mathcal{F}$ into a pair of Majorana fermions. However, one can already see that the mass-splitting will be very small. The sub-loop in Figure~\ref{fig:F_maj_mass} is the same loop-diagram that generates SM neutrino masses in Figure~\ref{fig:yukawa_loop}. Thus, in the limit that SM neutrino masses vanish, $m_\nu\rightarrow0$, the Majorana mass for $\mathcal{F}$ will also vanish. We therefore expect $\Delta M_{\mathcal{F}}\propto m_\nu$, where  $\Delta M_{\mathcal{F}}$ is the Majorana mass for $\mathcal{F}$. This is born out by explicit calculations. For example, with $M_{\mathcal{F}}\ll M_S$, where $M_S$ denotes an approximate common mass for the beyond-SM scalars, one obtains
\bea
\Delta M_{\mathcal{F}}&\sim &\frac{\lambda_1^2\lambda_{11\h}}{16\pi^2}\, \frac{\langle H\rangle^2}{M_S^2}\times m_\nu ,
\eea
where the Lagrangian contains the term $\lambda_{11\h}^*(H^\dagger S_1)^2\subset\mathcal{L}$ to generate the uppermost vertex in Figure~\ref{fig:F_maj_mass}. The beyond-SM neutral fermions therefore form pseudo-Dirac particles with a tiny splitting.

Direct detection experiments give strong constraints on spin-independent elastic-scattering events that can occur when fermionic dark-matter couples to the $Z$ boson~\cite{Aprile:2011hi}.  These constraints can be avoided if the Dirac fermion has a mass-split, as the resulting pair of Majorana fermions has non-diagonal couplings to the $Z$ boson (to leading order). Provided the mass split exceeds the average kinetic-energy of the local dark-matter particles, $Z$-boson exchange with SM detectors is highly suppressed, as the heavier fermion is kinematically inaccessible. However, for the models with neutral fermions in Table~\ref{L_yukawa} the mass split satisfies
\bea
\frac{\Delta M_{\mathcal{F}}}{M_{\mathcal{F}}}& <&\frac{m_\nu}{M_{\mathcal{F}}} \ \ \lesssim\ \  10^{-12}\quad\quad \mathrm{for}\quad M_{\mathcal{F}}=\mathcal{O}(\mathrm{TeV}),
\eea
which is (much) too small to evade direct-detection bounds, given typical DM speeds of $v_{\dm}\sim10^{-3}$. We conclude that none of the models in Table~\ref{L_yukawa} are viable when the lightest beyond-SM field is a fermion, due to either a cosmologically excluded stable charged-particle or a dark-matter candidate that contradicts direct-detection constraints. The entire region of parameter space in which a fermion is the lightest beyond-SM state is thus excluded for these models.
\section{Inert Scalar Dark Matter\label{sec:scalar_dm}}
With the above information we can restrict our attention to the limit $M_{\mathcal{F}}\gg M_{1,2}$ for the models in Table~\ref{L_yukawa}, for which the stable particle is a scalar. In this limit the models are effectively inert $N$-tuplet models with the additional feature of realizing radiative neutrino mass. In this section, we consider the viability of the scalar dark-matter candidates in the different models.

We first consider models $(A)$ and $(B)$, whose common features allow them to be  discussed together. Both these models have a single (candidate) dark-matter multiplet, which is an inert SM-like scalar doublet, $S_1\sim(1,2,1)$; i.e.~an inert doublet~\cite{Barbieri:2006dq}. Also, in both models the second scalar is a charged doublet, $S_2=(S_2^{++},S_2^+)^T$, whose components  must be heavier than the dark matter.  Inert-doublet dark matter is well-studied in the literature, and it is known that a viable dark-matter abundance can be realized~\cite{Barbieri:2006dq}. The inert-doublet leads to three new scalars, which we denote as $H'^{\pm}$, $H'^0$ and $A^0$, and either of the last two can be the dark matter. As per usual for an inert-doublet model, the neutral components of $S_1$ cannot mix with the SM Higgs in models $(A)$ and $(B)$, due to the discrete symmetry. However, the charged scalar $S_1^+$ will mix with $S_2^+$, as mentioned already. If this mixing is large, the phenomenology of the lightest charged-scalar will differ from that of $H'^+$ in a standard inert-doublet model. For small mixing the lightest charged scalar will correspond mostly to $H'^{+}$ and the phenomenology of $S_1$ will be well approximated by a standard inert-doublet. Note that one cannot take the limit $\lambda_{\sh}\rightarrow0$ without turning off the radiative neutrino mass in Figure~\ref{fig:yukawa_loop}. The demand that radiative neutrino mass is realized therefore requires nonzero mixing between $S_1^+$ and $S_2^+$. However, given that $S_2$ must be heavier than the dark-matter,  one generally expects the mixing to be of order $\langle H^0\rangle^2 /M_2^2$ which is $\lesssim 10^{-1}$ for $M_2\gtrsim$~TeV. Thus, $S_1$ can be well-approximated by a standard inert-doublet.

The inert-doublet model contains five main regions of parameter space in which the observed relic-abundance is obtained~\cite{Dolle:2009fn}. Four of these have a light particle-spectrum that can be probed at the LHC.  The discovery of an SM-like scalar with mass of roughly $125$~GeV at the LHC allows one to update the viable parameter space and phenomenology of the inert-doublet models.  Recent analysis, incorporating the LHC data, shows that the low-mass regions for the dark-matter candidate can already be in tension with constraints from XENON100~\cite{Aprile:2011hi} and WMAP~\cite{Komatsu:2010fb}, while the heavier region with $M_{\dm}\gtrsim500$~GeV is essentially unaffected~\cite{Goudelis:2013uca}. Specifically, the surviving region for lighter dark-matter lies close to the Higgs-resonance/$WW$-production threshold~\cite{Goudelis:2013uca}. 

The region of parameter space with $M_{\dm}=\mathcal{O}(10)$~GeV is particularly interesting for the present models as, in this case, the additional beyond-SM multiplets can be light enough to appear at the LHC. This was discussed already in Ref.~\cite{Aoki:2011yk} for model $(A)$, where it was shown that the charged scalar-doublet $S_2\sim(1,2,3)$ can produce observable signals when the inert-doublet dark matter is light. Although the region of parameter space with heavier dark-matter ($M_{\dm}\gtrsim500$~GeV) will not be accessible at the LHC, it is expected that XENON-1T will probe this parameter space, potentially giving observable direct-detection signals~\cite{Klasen:2013btp}. One deduces that viable dark matter is possible in both models $(A)$ and $(B)$, and that the combined (projected)  LHC and XENON-1T data sets are expected to probe the viable parameter space in these models.

We note that, in general, models with hypercharge-less dark matter are not as strongly constrained by direct-detection experiments. Such candidates do not couple directly to the $Z$ boson so interactions with detectors do not arise at tree-level. Provided the mass-splitting between the charged  and neutral components of the dark-matter multiplet exceeds the average kinetic-energy of the dark matter in the local halo, interactions with the $W$ boson are also highly suppressed (or absent). Even if the neutral and charged components of a dark-matter multiplet are degenerate at tree-level, an $\mathcal{O}(100)$~MeV split is induced radiatively, which is sufficient to ensure scattering via $W$ boson exchange is suppressed/absent.

These comments apply to model $(D)$, in which the sole dark-matter candidate is the neutral component of the inert real-triplet, $S_1\sim(1,3,0)$. The neutral component of this multiplet does not interact with the $Z$ boson, and the charged component can be sufficiently split  (by radiative effects) to ensure the neutral state is the lightest field. This alleviates potential tension with direct-detection experiments. The possibility of inert real-triplet dark matter is well known in the literature~\cite{Cirelli:2005uq,FileviezPerez:2008bj,Hambye:2009pw,Araki:2011hm}. The neutral component of $S_1$ is a viable cold dark-matter candidate that saturates the observed relic abundance of $\Omega_{\cdm}\hat{h}^2\simeq0.11$~\cite{Komatsu:2010fb} for $M_{\dm}\approx 2.5$~TeV~\cite{Cirelli:2005uq}. If the real-triplet is lighter it can only comprise part of the dark-matter abundance and additional candidates are needed. In model $(D)$ the dark-matter abundance must be comprised solely of $S_1$, so that $M_{\dm}\approx 2.5$~TeV is a necessary requirement for this model. Unfortunately this makes it difficult to directly produce the exotic states at the LHC; both $\mathcal{F}\sim(1,2,-1)$ and $S_2\sim(1,1,2)$ must be heavier than $2.5$~TeV to ensure the dark matter is the lightest exotic, pushing them beyond projected experimental reach. Model $(D)$ is, however,  a viable model of dark matter and radiative neutrino mass for $M_{\dm}\approx 2.5$~TeV.

Next we turn our attention to models $(F)$ and $(G)$, which both employ $\mathcal{F}\sim(1,2,-3)$ and $S_1\sim(1,3,2)$. In both models $S_2$ is comprised purely of electrically-charged fields, so the neutral component of $S_1$ is the sole dark-matter candidate. This complex neutral-field cannot mix with the SM scalar, due to the $Z_2$ symmetry, so its particle and antiparticle sates remain degenerate. The dark-matter abundance is therefore comprised of both states, posing a serious difficulty for these models. Due to the nonzero hypercharge for $S_1$, the neutral field can scatter off SM-detectors via tree-level $Z$-boson exchange. This process is strongly constrained by direct-detection data sets. Previous works show that one requires a mass of $\sim 2.6$~TeV to obtain the correct abundance, however, the spin-independent cross-section  exceeds $10^{-37}$cm$^2$ in the regions of parameter space compatible with the LEP experiments~\cite{Hambye:2009pw,Araki:2011hm}. Such a large cross section is incompatible with the constraints from, e.g.,~XENON100~\cite{Aprile:2011hi}. Thus, although one can successfully generate the requisite dark-matter abundance, direct-detection constraints prove fatal for models $(F)$ and $(G)$ and both models can be excluded. 

It remains for us to consider models $(C)$ and $(E)$. These models admit two distinct scalar dark-matter candidates and thus allow more possibilities, as we shall see in following section. 
\section{Models with a Real Scalar and a Complex Triplet\label{sec:real_and_complex_scalar}}
Models $(C)$ and $(E)$ both employ the SM-like fermion $\mathcal{F}\sim(1,2,-1)$ and the complex scalar triplet  $S_2\sim(1,3,2)$. Furthermore, in both cases $S_1$ is a real scalar. These models differ from the other cases as both scalars now possess a neutral component, giving two dark-matter candidates. We saw that models $(F)$ and $(G)$ could be excluded precisely because the dark-matter abundance was comprised of the neutral component of the complex scalar triplet. This difficulty is avoided in models $(C)$ and $(E)$, however, due to allowed mass-mixing between the  neutral components of $S_1$ and $S_2$. In this section we discuss model $(C)$ in some detail, to elucidate the possibilities. The analysis of model $(E)$ is rather similar, due to the related field content, and we limit ourselves to  some brief comments on this model at the end of the section.

Model $(C)$ contains the beyond-SM scalars $S_1\sim(1,1,0)\equiv S$ and $S_2\sim(1,3,2)\equiv \Delta$. The full scalar-potential can be written as
\bea
V(H,S,\Delta)&=&\frac{-\mu^2}{2}|H|^2 +\frac{\tilde{M}_S^2}{2}S^2 +\tilde{M}_\Delta^2 \mathrm{Tr}(\Delta^\dagger \Delta) +\lambda_1 |H|^4 +\lambda_2\,[\mathrm{Tr}(\Delta^\dagger \Delta) ]^2\nonumber\\
& &+\lambda_2'\,\mathrm{Tr}(\Delta^\dagger \Delta\Delta^\dagger \Delta) +\frac{\lambda_3}{2}S^4+\lambda_4\, |H|^2\,\mathrm{Tr}(\Delta^\dagger \Delta) + \lambda_4' \,H^\dagger \Delta^\dagger\Delta H\nonumber\\
& &+\lambda_5|H|^2S^2 +\lambda_6\, S^2\,\mathrm{Tr}(\Delta^\dagger \Delta)+\lambda_{\sh} S\left\{ \tilde{H}^\dagger \Delta^\dagger H +H^\dagger \Delta \tilde{H}\right\},\label{model_c_potential}
\eea
where the overall phase of $\Delta$ has been used to choose $\lambda_{\sh}$ real without loss of generality.\footnote{A related potential was considered in Ref.~\cite{Foot:2007ay}.} The discrete symmetry $\{S,\,\Delta\}\rightarrow -\{S,\,\Delta\}$ ensures there is no mass-mixing between the SM scalar and the beyond-SM fields. This symmetry also forbids terms linear in a single beyond-SM scalar, like $H\Delta^\dagger H$ or $SH^2$, which would otherwise induce a non-zero VEV for $\Delta$ and $S$ after electroweak symmetry breaking. Consequently, parameter space exists in which neither $S$ nor $\Delta$ acquire a VEV. The scalar $S$ and the neutral components of $\Delta$ will mass-mix, however, due to the $\lambda_{\sh}$ term in the potential. 

Expanding the neutral SM-scalar around its VEV, and expanding the neutral component of $\Delta$ as
\bea
H^0= \frac{1}{\sqrt{2}}(v+h^0+i\chi^0) \quad\mathrm{and}\quad \Delta^0=\frac{1}{\sqrt{2}}(\Delta_R+i\Delta_I),
\eea
respectively, the mass-mixing Lagrangian for the neutral scalars is
\bea
\mathcal{L}&\supset& -\frac{1}{2} \mathcal{S}^T\mathcal{M}^2 \mathcal{S}.
\eea
Here the basis vector is $\mathcal{S}=(S, \Delta_R,\Delta_I)^T$, and the squared-mass matrix has the form
\bea
\mathcal{M}^2 = 
\left(
\begin{array}{ccc}
\tilde{M}_S^2 +\lambda_5 v^2& \frac{\lambda_{\sh}}{2\sqrt{2}}v^2&0\\
 \frac{\lambda_{\sh}}{2\sqrt{2}}v^2& \tilde{M}_\Delta^2 +\frac{\lambda_4}{2}v^2&0\\
0&0& \tilde{M}_\Delta^2 +\frac{\lambda_4}{2}v^2
\end{array}
\right).
\eea
Thus, the CP-odd scalar $\Delta_I$ is a mass eigenstate with mass $\tilde{M}_{\Delta}^2+\lambda_4 v^2/2$, while the CP-even scalars $S$ and $\Delta_R$ mass-mix to produce two physical scalars that are linear combinations of these fields. 

The dark-matter candidate will be one of the neutral-scalar mass eigenstates. To determine which one, we must find the masses for the mixed CP-even states. Let us define $M_S^2=\tilde{M}_S^2 +\lambda_5 v^2$ and $M^2_\Delta= \tilde{M}_\Delta^2 +\lambda_4v^2/2$, which are the CP-even mass eigenstates in the limit $\lambda_{\sh}\rightarrow 0$.  In this limit $\Delta_R$ and $\Delta_I$ are degenerate and form a single complex-scalar with mass $M^2_\Delta$. For nonzero $\lambda_{\sh}$, the CP-even mass eigenvalues can be written as
\bea
M_{\pm} = \frac{1}{2}\left\{M_S^2+M^2_\Delta \pm\left[ (M_S^2-M^2_\Delta)^2 + \frac{\lambda_{\sh}^2}{2}v^4\right]^{1/2}\right\},
\eea
where the eigenstates are related to the original fields as
\bea
\left(
\begin{array}{c}
S_+\\
S_-
\end{array}
\right)=
\left(
\begin{array}{cc}
\cos\theta&\sin\theta\\
-\sin\theta&\cos\theta
\end{array}
\right)\,
\left(
\begin{array}{c}
S\\
\Delta_R
\end{array}
\right).
\eea
Here, the mixing angle is
\bea
\tan2\theta&=&\frac{\lambda_{\sh}v^2}{\sqrt{2}(M^2_S-M^2_\Delta)}. 
\eea

In the limit where the singlet-scalar is heaviest,  $M_S^2\gg M^2_\Delta$, the mass eigenvalues are approximately
\bea
M_+^2\simeq M_S^2+\frac{\lambda_{\sh}^2v^4}{8M_S^2}\quad\mathrm{and}\quad M_-^2\simeq M^2_\Delta - \frac{\lambda_{\sh}^2v^4}{8M_S^2}\quad\mathrm{for}\quad M_S^2\gg M^2_\Delta.
\eea
Noting that  $M_-^2<M^2_\Delta$, reveals that $S_-$ is the lightest exotic state and is thus the DM candidate. Simple expressions are obtained for the mass eigenvectors in this limit:
\bea
\left.
\begin{array}{c}
S_+\simeq S+\frac{\lambda_{\sh} v^2}{2\sqrt{2} M_S^2}\Delta_R\\
\\
S_-\simeq \Delta_R-\frac{\lambda_{\sh} v^2}{2\sqrt{2} M_S^2}S
\end{array}
\right.
\quad\quad\mathrm{for}\quad M_S^2\gg M^2_\Delta,
\eea 
so that the lightest scalar $S_-$ is comprised mostly of $\Delta_R$. 

Thus, for $M_S^2\gg M^2_\Delta$ the dark matter is comprised of $S_-$, which mostly consists of the CP-even part ($\Delta_R$) of the neutral field in the  scalar triplet $S_2$. The mass-splitting between $S_-$ and the CP-odd state $\Delta_I$ is $|\Delta M^2|= \lambda_{\sh}^2v^4/8M_S^2$. Provided this splitting exceeds the dark-matter kinetic-energy,  $\sqrt{|\Delta M^2|}> \mathrm{KE}_{\dm}$,  the state $\Delta_I$ will not be kinematically accessible via tree-level processes in direct-detection experiments. This significantly weakens the bounds on dark matter arising from a complex triplet. This also gives an upper bound on the mass parameter $M_S^2$, beyond which the splitting between the dark matter and $\Delta_I$ is so small that tree-level scattering via $Z$ exchange is expected in present-day experiments. One finds
\bea
M_S &<& \frac{1}{2\sqrt{2}}\,\frac{|\lambda_{\sh}|\, v^2}{\mathrm{KE}_{\dm}}\ \simeq \ \left(\frac{|\lambda_{\sh}|}{10^{-2}}\right)\, \left(\frac{10^{-3}}{v_{\dm}}\right)^2\left(\frac{2.6~\mathrm{TeV}}{M_{\dm}}\right)\times 10^2~\mathrm{TeV}.
\eea
Thus, the heavier state $S_+$ cannot be made arbitrarily heavy if the dark matter is to avoid exclusion via, e.g., the XENON100 data.

 With $M_{\dm}\equiv M_-=\mathcal{O}(\mathrm{TeV})$, the mass-split between $S_-$ and $\Delta_I$ is smaller than $M_{\Delta}$ (the mass of $\Delta_I$). Once the temperature drops below $M_\Delta$, the heavier state $\Delta_I$ will decay, with the decay products necessarily  containing $S_-$, due to the conserved discrete-symmetry.  The expression $|\Delta M^2|= \lambda_{\sh}^2v^4/8M_S^2$ shows that the mass-splitting between $S_-$ and $\Delta_I$ is bounded as $\sqrt{|\Delta M^2|}\lesssim |\lambda_{\sh}|\times 4$~GeV, given that $M_{\Delta}\gtrsim3$~TeV is needed to achieve the correct relic abundance and we are working with $M_S>M_\Delta$. In this mass range $\Delta_I$ can decay as $\Delta_I\rightarrow S_-+Z^*\rightarrow S_-+\bar{f}f$, where $f$ is a SM fermion with mass $m_f<|\lambda_{\sh}|\times2$~GeV. Therefore, even if charged SM fermions are not kinematically available,  final-states containing neutrinos will be accessible unless $\lambda_{\sh}$ is exceptionally small. After $\Delta_I$ has decayed away, the primordial plasma is comprised of $S_-$ and the SM fields. $S_-$ can maintain equilibrium with the SM sector via gauge interactions and via the $\lambda_4$ and $\lambda_4'$ (hereafter $\lambda_4$) quartic terms in Eq.~\eqref{model_c_potential}.\footnote{$S_-$ also interacts with the SM via the $\lambda_5$ and $\lambda_{\sh}$ terms in Eq.~\eqref{model_c_potential}. However, there is only a small admixture of $S$ in $S_-$  for the limit we discuss, so these interactions are suppressed by the small mixing angle $\theta= \mathcal{O}(\lambda_{\sh} v^2/M_S^2)\ll1$.} When the quartic interactions are dominant, the model is similar to an inert real-triplet model; the dark-matter abundance will be obtained for $M_{\dm}\simeq 2.5$~TeV, in line with the analysis of Ref.~\cite{Cirelli:2005uq}. As one makes $\lambda_4$ smaller, the tree-level mass splitting between the charged and neutral components of $\Delta$ diminishes and coannihilation channels like $\Delta^-\Delta^{++}\rightarrow W^+\gamma$ become available. At this point, making $\lambda_4$ smaller does not modify the requisite dark-matter mass as gauge interactions dominate. The analysis of Ref.~\cite{Araki:2011hm} for an inert complex-triplet finds that $M_{\dm}\gtrsim 2.8$~TeV is required for the entire region of parameter space.\footnote{The exception being for the small resonant-region with $M_{\dm}\approx m_h/2$, for which the dark-matter mass is smaller due to the enhanced annihilation cross section.  We do not consider the resonant regions, which require tuned mass relations.} We thus expect that $M_{\dm}\gtrsim 2.5$~TeV will be required even when the gauge interactions dominate the quartic interactions during freeze-out  for the present scenario. With this value fixed, model $(C)$ becomes a viable model of neutrino mass and dark matter.  It will be difficult, however, to produce the exotics in this model, given that the lightest exotic mass is $\gtrsim 2.5$~TeV.\footnote{For a recent model with radiative neutrino mass that utilizes the resonant region to obtain viable dark matter from a complex scalar triplet, see Ref.~\cite{Kajiyama:2013zla}. If the dark-matter mass was similarly taken in the resonant region in our models, we could also take the extra exotics to be $\mathcal{O}(100)$~GeV. Also see Ref.~\cite{Kanemura:2012rj}.} 

We now briefly discuss the alternative limit with $M_S^2\ll M^2_\Delta$. In this case the CP-even mass eigenvalues are
\bea
M_+^2\simeq M_\Delta^2+\frac{\lambda_{\sh}^2v^4}{8M_\Delta^2}\quad\mathrm{and}\quad M_-^2\simeq M^2_S - \frac{\lambda_{\sh}^2v^4}{8M_\Delta^2}\quad\mathrm{for}\quad M_\Delta^2\gg M^2_S.
\eea
We see that the dark-matter candidate remains as the lightest CP-even eigenstate $S_-$, with mass $M_-$. The mass eigenstates are now 
\bea
\left.
\begin{array}{c}
S_+\simeq \Delta_R+\frac{\lambda_{\sh} v^2}{2\sqrt{2} M_\Delta^2}S\\
\\
S_-\simeq S-\frac{\lambda_{\sh} v^2}{2\sqrt{2} M_\Delta^2}\Delta_R
\end{array}
\right.
\quad\quad\mathrm{for}\quad M_\Delta^2\gg M^2_S,
\eea
so the dark matter is comprised mostly of the singlet-scalar $S$. Singlet-scalar dark matter is well known~\cite{McDonald:1993ex} and detailed analysis show that $M_{\dm}\gtrsim 80$~GeV is compatible with direct-detection constraints and WMAP data for a Higgs mass of $m_h\simeq125$~GeV~\cite{Djouadi:2011aa}. The viable region of parameter space can be probed by XENON1T, excepting a small resonant window with $M_{\dm}\simeq62$~GeV, where the dark-matter-Higgs coupling can be very small. Lighter dark-matter with $M_{\dm}\lesssim60$~GeV is ruled out by LHC bounds on invisible Higgs decays~\cite{Djouadi:2011aa}.

We see that model $(C)$ has viable parameter space in which it  behaves like an inert-triplet model or an inert singlet model. This analysis is sufficient to demonstrate that model $(E)$ also has viable regions of parameter space. In model $(E)$ one has $S_1\sim(1,3,0)$, while the second scalar remains as $S_2\sim(1,3,2)$. The scalar potential for this model contains a term similar to the $\lambda_{\sh}$ term in Eq.~\eqref{model_c_potential}, which mixes the neutral component of $S_1$ with the CP-even neutral component of $S_2$. If $S_2$ is heaviest the model behaves like an inert real-triplet model, while if $S_1$ is heaviest the lightest scalar is mostly comprised of $\Delta_R$ (the neutral CP-even part of $S_2$). Direct-detection constraints can be evaded due to the mass mixing, and the model is again an effective model of inert-triplet dark matter. In both cases we expect that a viable dark-matter abundance and viable neutrino masses can be obtained, though the dark matter will be heavy, with $M_{\dm}\gtrsim2.5$~TeV (neglecting resonant regions).

In terms of the observational prospects for the beyond-SM multiplets at the LHC, the limit $M_\Delta^2\gg M^2_S$ in model $(C)$ appears to be the most optimistic scenario for the models in Table~\ref{L_yukawa} (excepting the resonant regions, which also allow lighter fields). In this limit the dark matter can be relatively light, $M_{\dm}\simeq100$~GeV, and thus the exotic states $\Delta$ and $\mathcal{F}$ can both be of order a few hundred GeV.  In principle it could be possible to observe all three beyond-SM multiplets in this limit. For the other viable models in Table~\ref{L_yukawa} the dark matter has to be relatively heavy: $M_{\dm}\gtrsim 500$~GeV for the inert doublet models, and $M_{\dm}\gtrsim 2.5$~TeV for the inert triplet cases. This pushes the additional beyond-SM fields beyond the reach of the LHC.


Note that any mass degeneracy between charged and neutral members of an inert multiplet at tree-level is lifted by radiative effects, making the charged components heavier than the neutral components. The heavier members of a given multiplet can decay to lighter members of the same multiplet via the weak interactions, e.g.~$\mathcal{F}^+\rightarrow W^++\mathcal{F}^0$, where the $W$ can be virtual.  A heavier multiplet can also decay to a lighter multiplet via the Yukawa coupling; e.g.~$\mathcal{F}^-\rightarrow S^0 + \ell^-$ if $M_{\mathcal{F}}\gg M_S$. Because of the discrete symmetry the new fields can only be pair produced in colliders, and conservation of the $Z_2$ charge means final states resulting from exotic decay chains necessarily include stable electrically-neutral fields that will escape the detector.


\section{On the Origin of the Discrete Symmetry\label{sec:gauge_symmetry}}
Following Ma's original proposal, we employed a discrete $Z_2$ symmetry to ensure stability of the lightest beyond-SM field appearing in the neutrino-mass diagram. One can argue that the use of a discrete symmetry is not completely satisfying, either because it seems ad~hoc, or because of the view that quantum gravity effects are not expected to conserve global symmetries. This motivates one to consider whether a simple explanation for the discrete symmetry can be found. The simplest possibility is to replace the discrete symmetry with a gauged $U^\prime(1)$ symmetry, which would not be broken by quantum gravity effects. With enough additional ingredients one can presumably achieve this goal for all the models we have discussed. However, we would like to know which models allow for a minimal extension, such that $Z_2\rightarrow U^\prime(1)$, and a single SM-singlet scalar $\eta$ is added to the particle spectrum to break the $U^\prime(1)$ symmetry.

Writing  the full gauge group as $\mathcal{G}_{\sm}\times U(1)^\prime$, where $\mathcal{G}_{\sm}$ is the SM gauge group, we have the following transformation properties for the beyond-SM fields\footnote{Note that real scalars must be taken as complex once they are given a nonzero $U^\prime(1)$ charge.}
\bea
\eta \sim(1_{\sm},Q_\eta)\,, \quad S_{1,2}\sim( Q^{\sm,}_{1,2},Q)\,,\quad \mathcal{F}\sim(Q^{\sm}_{\mathcal{F}},-Q),
\eea
where the ``SM" superscript denotes the charges under $\mathcal{G}_{\sm}$, given in Table~\ref{L_yukawa}. Inspection of Eq.~\eqref{eq:L_ykawa_lagrange} shows that all Lagrangian terms needed to generate neutrino mass are allowed by the $U^\prime(1)$ symmetry. However, in the case of models $(A)$ and $(B)$, which are inert-doublet models, the enhanced symmetry prevents the additional term $(S_1^\dagger H)^2$. This term is not needed for neutrino mass but is required to split the neutral components of $S_1\sim(1,2,1)$ in order to avoid direct-detection constraints~\cite{Barbieri:2006dq}. Thus, models $(A)$ and $(B)$ are not compatible with this minimal symmetry extension. 

On the other hand, we find that models $(C)$, $(D)$ and $(E)$, which have one scalar forming a real representation of the SM gauge symmetry, remain as viable models of dark matter provided $Q_\eta=-2Q$. This relationship is needed to lift a mass-degeneracy of neutral beyond-SM fields. For example, consider model $(C)$, which now has the following terms in the scalar potential
\bea
V(H,S,\Delta,\eta)&\supset&\lambda_{\sh} \left\{ S\tilde{H}^\dagger \Delta^\dagger H +S^*H^\dagger \Delta \tilde{H}\right\}\ +\ \frac{\mu_{\eta }}{2}\left\{S^2\eta+ S^{*2}\eta^\dagger\right\},\label{model_c_potential'}
\eea
in addition to the terms in Eq.~\eqref{model_c_potential}. All other terms containing $\eta$ depend only on the modulus $|\eta|^2$. We have used the relative phase of $S$ and $\Delta$ to choose $\lambda_{\sh}$ real and the phase of $S$ to choose $\mu_{\eta }$ real. Note that the symmetry breaking $U^\prime(1)\rightarrow Z_2$ is achieved by nonzero $\langle \eta\rangle$, motivating the discrete symmetry as an accidental subgroup of the gauged $U^\prime(1)$ symmetry.

In the basis $\mathcal{S}=(S_R, \Delta_R,S_I,\Delta_I)^T$, the squared-mass matrix has the form\footnote{In making $S$ complex, some of the coupling/mass parameters must be scaled, relative to the real-$S$ case, to obtain this form.}
\bea
\mathcal{M}^2 = 
\left(
\begin{array}{cccc}
\tilde{M}_S^2 +\lambda_5 v^2+2\mu_{\eta } \langle\eta\rangle& \frac{\lambda_{\sh}}{4}v^2&0&0\\
 \frac{\lambda_{\sh}}{4}v^2& \tilde{M}_\Delta^2 +\frac{\lambda_4}{2}v^2&0&0\\
0&0& \tilde{M}_S^2 +\lambda_5 v^2-2\mu_{\eta } \langle\eta\rangle& \frac{\lambda_{\sh}}{4}v^2\\
0&0& \frac{\lambda_{\sh}}{4}v^2& \tilde{M}_\Delta^2 +\frac{\lambda_4}{2}v^2
\end{array}
\right).
\eea
Observe that the entries for the CP-even and CP-odd states are identical in the limit $\mu_\eta\rightarrow0$. This would produce degenerate states so that, in the case where the dark matter is comprised mostly of $\Delta$, the dark matter would be ruled out by XENON100 (it would be an inert complex-triplet). For non-zero $\mu_\eta$, however, the CP-even and CP-odd states are non-degenerate and viable dark-matter is achieved. When the dark matter is mostly (or completely, for model $(D)$) comprised of a real representation of $\mathcal{G}_{\sm}$, the splitting achieved by nonzero $\mu_\eta$ also ensures direct detection signals resulting from mixing between $Z^\prime$ and $Z$ are suppressed.\footnote{The interactions with $Z^\prime$ are non-diagonal in the real and imaginary components of the beyond-SM scalars; the splitting means one of these states is not part of the present-day dark-matter abundance, thereby suppressing scattering via $Z^\prime$ exchange. Note that the Lagrangian contains a kinetic-mixing term between $U^\prime(1)$ and SM hypercharge, but even if one takes the relevant coupling to be small, mixing between $Z'$ and $Z$ is induced by loops containing exotics charged under both $U^\prime(1)$ and $\mathcal{G}_{\sm}$.} 

There will be an additional scattering process for the dark matter due to the mixing between $\eta$ and the SM scalar, which gives a standard Higgs portal interaction. Given that the coupling for this interaction is not needed to achieve the observed dark-matter abundance, one can always choose this coupling to be small enough to comply with constraints. Thus, with this simple gauge extension, we can explain the origin of the discrete symmetry for models $(C)$, $(D)$ and $(E)$, while retaining the desirable features of radiative neutrino mass and a viable dark-matter abundance.

Note that Ma's original proposal~\cite{Ma:2006km}, and the variant using a real triplet fermion~\cite{Ma:2008cu}, are not compatible with this minimal symmetry upgrade; in the case of scalar dark-matter, the $(S^\dagger H)^2$ term is precluded, meaning direct-detection experiments rule the model out, similar to the minimal gauge extension of models $(A)$ and $(B)$. Furthermore, one  encounters gauge-anomalies given that $\mathcal{F}_R$ is a chiral field in Refs.~\cite{Ma:2006km,Ma:2008cu} --- additional model building is therefore needed to explain the origin of the $Z_2$ symmetry in these cases. We do not pursue this matter here.
\section{Beyond the Adjoint Representation\label{sec:non-minimal_models}}
In the preceding sections we studied generalizations of Ma's 2006 model with radiative neutrino mass and stable dark-matter candidates. In doing so we restricted our attention to beyond-SM multiplets no larger than the adjoint representation. As mentioned already, one can generate neutrino mass via Figure~\ref{fig:yukawa_loop} and obtain dark-matter candidates in models with larger multiplets. We briefly discuss this matter in the present section.

First consider the case with a Majorana mass insertion, as in Figure~\ref{fig:mass_loop}. Allowing for $SU(2)$ representations as large as the quintuplet-rep. we  find  two additional models. Both of these  employ the quadruplet scalar $S\sim(1,4,1)$,  with the real fermion being either a triplet $\mathcal{F}\sim(1,3,0)$, or a quintuplet $\mathcal{F}\sim(1,5,0)$. The latter model was detailed in Ref.~\cite{Kumericki:2012bh}.\footnote{The real quintuplet fermion has been studied as a ``Minimal Dark-Matter" candidate, due to an accidental symmetry that arises when one adds this field to the SM~\cite{Cirelli:2009uv}. However, this symmetry is broken when an additional scalar is added to allow radiative neutrino masses; a discrete symmetry must be imposed to retain the dark-matter candidate.} In both models we expect either scalar or fermionic dark-matter is possible,  as in Ma's original proposal; the neutral fermion does not couple to the $Z$ boson and can therefore remain consistent with direct-detection constraints.

Generalizing the models with a fermion mass insertion of the Dirac type (i.e.~generalizing the models in Table~\ref{L_yukawa}), we find more variants are possible. For completeness we list these in the Appendix, but here offer the following comments. As with the models in Table~\ref{L_yukawa}, we find that fermionic dark-matter can be ruled out for all models with larger gauge representations. The fermions remain as pseudo-Dirac particles with tiny splittings, set by the SM neutrino masses. Such small splittings permit unsuppressed tree-level scattering with SM detectors via $Z$-boson exchange, which is ruled out by XENON100. We thus rule out the parameter space in which the fermion is the lightest beyond-SM state, for the same reasons as discussed in Section~\ref{sec:fermion_dm}. For the case of scalar dark-matter, one has to consider the individual models, as was needed for the models in Table~\ref{L_yukawa}. Some models can be immediately ruled out for the same reasons that models $(F)$ and $(G)$ could be excluded; for example, model $(L)$ in Table~\ref{L_yukawa_quadruplet} can be excluded as it gives an inert complex-triplet model. Similarly model $(R)$ in Table~\ref{L_yukawa_quintuplet}  is ruled out, as $S_2$ contains only charged components and $S_1$ has nonzero hypercharge. The other models appear to be compatible with direct-detection constraints, provided the neutral components of the scalars mix when the lightest scalar has nonzero hypercharge, much as models $(C)$ and $(E)$ were viable. For example, model $(M)$ contains $S_1\sim(1,4,1)$ as the only beyond-SM scalar with a neutral component. However, the Lagrangian allows a term $\lambda (S_1^\dagger H)^2\subset\mathcal{L}$ that can split the components of the neutral scalar, allowing one to avoid direct-detection constraints (this is analogous to the splitting obtained in an inert-doublet model).

Finally, we note that the use of larger multiplets may have an additional phenomenological benefit. Ref.~\cite{Kopp:2013mi} shows that large multiplets that mediate interactions between dark matter and the SM can enhance loop-induced annihilation of dark matter into $2\gamma$ and $\gamma+Z$ final states, without requiring non-perturbatively large couplings. This occurs because the larger multiplets admit fields with larger electric-charges, naturally enhancing loop-processes with final-state photons. It does not appear to be possible to realize the astrophysical gamma-ray signal~\cite{Weniger:2012tx} in the models presented in the Appendix, but simple extensions do seem compatible with this idea. For example, model $(N)$ in Table~\ref{L_yukawa_quintuplet} employs $\mathcal{F}\sim(1,4,-1)$, $S_1\sim(1,5,0)$ and $S_2\sim(1,5,2)$. When $S_1$ is the lightest beyond-SM state the dark matter is comprised (mostly) of the neutral component of $S_1$.  There is a one-loop contribution to processes like $DM+DM\rightarrow 2\gamma,\gamma+Z$, containing virtual $S_2$ states in the loop, that is enhanced by the presence of the multiply charged component in $S_2$. Note that the dark-matter mass is required to be either $M_{\dm}\sim130$~GeV or $M_{\dm}\sim144$~GeV, in order to generate the gamma-ray excess via dark matter annihilations into either $2\gamma$ or $\gamma+Z$ final states, respectively. However, dark-matter comprised of $S_1\sim(1,5,0)$ is expected to have a mass $M_{\dm}\gtrsim5$~TeV in order to achieve the correct relic abundance~\cite{Cirelli:2005uq}, which is too large to explain the astrophysical signal. If one adds a singlet scalar $S$, that is also odd under the $Z_2$ symmetry,  to the model, then the region of parameter space where $S$ is the dark matter is compatible with $M_{\dm}\sim130$~GeV or $M_{\dm}\sim144$~GeV. The components of $S_1$ and/or $S_2$ can then be $\mathcal{O}(100)$~GeV and the loop-processes advocated in Ref.~\cite{Kopp:2013mi} are present in the model, thereby enhancing the astrophysical gamma-ray signal.\footnote{Note that whether the $2\gamma$ final state or the $\gamma+Z$ final state is dominant, and thus whether the dark matter mass is $130$~GeV or $144$~GeV, depends on whether the dark matter $S$ couples  more strongly to $S_1$ or $S_2$. If the dominant coupling is to $S_2$, the $\gamma+Z$ final state is expected to dominate due to the nonzero hypercharge assignment for $S_2$, giving $M_{\dm}\sim144$~GeV. A dominant coupling to $S_1$ instead requires $M_{\dm}\sim130$~GeV as the $2\gamma$ final state is expected to dominate.} In this example there is a simple connection between the astrophysical signal and the mechanism of neutrino mass, with the large multiplets that enable the latter also enhancing the former. It could be interesting to take these ideas further to see if the dark matter can be realized as one of the fields in the neutrino loop-diagram, rather than an extra degree of freedom, or to study the phenomenology of the model just described.



\section{Conclusion\label{sec:conc}}
We studied a class of models with radiative neutrino mass and stable dark-matter candidates. Neutrino mass was generated by a one-loop diagram with the same topography as that proposed by Ma~\cite{Ma:2006km}. We generalized Ma's approach, detailing all variants with beyond-SM fields no larger than the adjoint representation. In the case where the neutrino mass diagram  contained a Majorana mass insertion, only two models were found, both of which were known. When the mass-insertion was of the Dirac type, such that lepton-number symmetry was broken by a vertex, we found a number of additional models. Fermionic dark-matter was excluded in all of these models, while two of the models were completely excluded due to direct-detection constraints. The remaining models allowed radiative neutrino mass and achieved a viable (scalar) dark-matter abundance. There were cases with an inert singlet, an inert doublet, and an inert triplet, providing a natural setting for inert $N$-tuplet models of dark-matter, with the additional feature of achieving radiative neutrino mass. Interestingly, some of the models allowed a simple extension, such that the (formerly imposed) discrete symmetry emerged as an accidental low-energy symmetry. We briefly discussed models with larger beyond-SM multiplets, showing that viable scenarios exist. With simple extensions,  the large multiplets enabling neutrino mass can also enhance present-day astrophysical gamma-ray signals, allowing a simple connection between the mechanism of neutrino mass and the astrophysical gamma-ray signal.
\section*{Acknowledgments\label{sec:ackn}}
The authors thank Y.~Kajiyama, K.~Nagao, H.~Okada, T.~Schwetz, A.~Strumia, and K.~Yagyu.
SSCL is supported in part by the NSC under Grant No.
NSC-101-2811-M-006-015 and in part by the NCTS of Taiwan. KM is supported by the Australian Research Council.
\appendix

\section{Mass Without the $\mathbf{Z_2}$ Symmetry\label{app:mass_no_Z2}}
Of the models presented in Table~\ref{L_yukawa}, only models $(A)$, $(B)$ and $(D)$ are expected to produce (dominant) radiative neutrino masses in the absence of the $Z_2$ symmetry. The other models contain the triplet scalar $S_{1,2}\sim(1,3,2)$, which Yukawa-couples to the SM leptons, and acquires a VEV due to the term $\mu HS_{1,2}H\subset V(H,S_{1},S_2)$, in the absence of the discrete symmetry. Thus, tree-level neutrino masses of the standard Type-II seesaw~\cite{type2_seesaw} form are expected to dominate the loop effect when the $Z_2$ symmetry is discarded.\footnote{The same is true for Ma's original proposal~\cite{Ma:2006km} and the triplet variant~\cite{Ma:2008cu}; if the $Z_2$ symmetry is turned off one obtains tree-level neutrino masses via a Type-I or Type-III seesaw, respectively.} For models $(A)$, $(B)$ and $(D)$, on the other hand,  tree-level neutrino masses do not arise if the $Z_2$ symmetry is removed, while the loop-diagram in Figure~\ref{fig:yukawa_loop} persists. Note that, if the $Z_2$ symmetry is turned off, the fermion $\mathcal{F}$ is not needed in order to generate nonzero radiative neutrino masses in model $(D)$~\cite{Law:2013dya}. However, the spectrum obtained without $\mathcal{F}$ is of the simplified-Zee form~\cite{Law:2013dya}, which is incompatible with the observed mixing pattern~\cite{He:2003ih}. Thus, the fermion $\mathcal{F}\sim(1,2,-1)$ is required to obtain a \emph{viable} mixing pattern in the absence of the $Z_2$ symmetry. 

Model $(B)$ has a similar particle content to the model presented in Ref.~\cite{Babu:2009aq}, modulo the replacement $S_2\sim(1,2,3)\rightarrow S_2\sim(1,4,3)$. This difference precludes the tree-level mass found in Ref.~\cite{Babu:2009aq} so model $(B)$ is purely a model of radiative masses, which could be studied without the discrete symmetry and dark matter.\footnote{One still requires a second SM-like doublet to achieve neutrino mass in this case, as the term $H^3S_2$ vanishes when there is only one SM doublet~\cite{Law:2013dya}. The SM scalar doublet would now appear inside the loop, however.}
\section{Models with Larger Multiplets\label{app:mass_non_minimal}}
In the text we found seven models that employ beyond-SM multiplets in either the fundamental or adjoint representation of $SU(2)_L$, and had an internal Dirac mass-insertion. In addition to these minimal models, one can realize radiative neutrino mass and dark-matter candidates with larger multiplets. We present the additional minimal models that arise if one permits multiplets forming the quadruplet  (isospin-$3/2$) representation of $SU(2)_L$ in Table~\ref{L_yukawa_quadruplet}. The labeling scheme follows on from Table~\ref{L_yukawa} in the text. If one allows for quintuplet multiplets there are additional models, shown in Table~\ref{L_yukawa_quintuplet} (also see Ref.~\cite{Picek:2009is} for a detailed example). The first case listed as model $(M)$ was presented in Ref.~\cite{McDonald:2013kca}.

\begin{table}
\centering
\begin{tabular}{|c|c|c|c|}\hline
& & &  \\
\ \ Model\ \ &
$\mathcal{F}$  & $S_1$&$S_2$\\
& & &  \\
\hline
& & &  \\
$(H)$& $\ \ (1,3,-2)\ \ $ &$\ \ (1,2,1)\ \ $&\ \ $(1,4,3)$\ \ \\ 
& & &  \\
\hline
& & &  \\
$(I)$& $\ (1,3,-2)\ $ &$\ (1,4,1)\ $&\ $(1,2,3)$\ \\ 
& & &  \\
\hline
& & &  \\
$(J)$& $\ (1,3,-2)\ $ &$\ (1,4,1)\ $&\ $(1,4,3)$\ \\ 
& & &  \\
\hline
& & &  \\
$(K)$& $\ (1,4,-1)\ $ &$\ (1,3,0)\ $&\ $(1,3,2)$\ \\ 
& & &  \\
\hline
& & &  \\
$(L)$& $\ (1,4,-3)\ $ &$\ (1,3,2)\ $&\ $(1,3,4)$\ \\ 
& & &  \\
\hline
\end{tabular}
\caption{\label{L_yukawa_quadruplet} Models with a Dirac mass insertion that employ quadruplet fields (Isospin-$3/2$).}
\end{table}

\begin{table}
\centering
\begin{tabular}{|c|c|c|c|}\hline
& & &  \\
\ \ Model\ \ &
$\mathcal{F}$  & $S_1$&$S_2$\\
& & &  \\
\hline
& & &  \\
$(M)$& $\ \ (1,4,-1)\ \ $ &$\ \ (1,4\mp1,0)\ \ $&\ \ $(1,4\pm1,2)$\ \ \\ 
& & &  \\
\hline
& & &  \\
$(N)$& $\ (1,4,-1)\ $ &$\ (1,5,0)\ $&\ $(1,5,2)$\ \\ 
& & &  \\
\hline
& & &  \\
$(O)$& $\ (1,4,-3)\ $ &$\ (1,4\mp1,2)\ $&\ $(1,4\pm1,4)$\ \\ 
& & &  \\
\hline
& & &  \\
$(P)$& $\ (1,4,-3)\ $ &$\ (1,5,2)\ $&\ $(1,5,4)$\ \\ 
& & &  \\
\hline
& & &  \\
$(Q)$& $\ (1,5,-2)\ $ &$\ (1,4,1)\ $&\ $(1,4,3)$\ \\ 
& & &  \\
\hline
& & &  \\
$(R)$& $\ (1,5,-4)\ $ &$\ (1,4,3)\ $&\ $(1,4,5)$\ \\ 
& & &  \\
\hline
\end{tabular}
\caption{\label{L_yukawa_quintuplet} Models with a Dirac mass insertion that employ quintuplet fields (Isospin-$2$).}
\end{table}

\end{document}